\documentclass[preprint,aps,draft,showpacs]{revtex4}
\begin{document}
\title{A kinetic approach to Bose-Einstein condensates: Self-phase modulation and Bogoliubov oscillations}
\author{J. T. Mendon\c{c}a$^a$}
\email{T.Mendonca@rl.ac.uk}
\author{R. Bingham$^b$}
\email{r.bingham@rl.ac.uk}
\affiliation{Rutherford Appleton Laboratory, Chilton, Didcot, Oxon OX11 OQX, U.K.}
\author{P. K. Shukla$^c$}
\email{ps@tp4.rub.de}
\affiliation{Institut f\"ur Theoretische Physik IV, Fakult\"at f\"ur Physik und Astronomie, 
Ruhr-Universit\"at Bochum, D-44780 Bochum, Germany}
\date{Received 8 November 2004}
\begin{abstract} A kinetic approach to the Bose-Einstein condensates (BECs) is proposed, based on the Wigner-Moyal equation (WME). In the 
quasi-classical limit, the WME reduces to the particle number conservation equation. 
Two examples of applications are: i) a self-phase modulation of a BE condensate 
beam where we show that a part of the beam is decelerated and eventually 
stops as a result of the gradient of the effective self-potential; ii) the derivation of a kinetic 
dispersion relation for sound waves in the BECs, including a collisionless Landau damping.

\bigskip
\noindent $^a$ On leave from the Instituto Superior T\'{e}cnico, 1049-001 Lisboa, Portugal.

\bigskip
\noindent $^b$ Also at the Department of Physics, University of Strathclyde, Glasgow G4 0NG, Scotland.

\bigskip
\noindent$^c$ Also at the Department of Physics, Ume\aa~University, SE90187 Ume\aa,~Sweden.

\end{abstract}

\maketitle

\newpage
\section{Introduction}

Presently, the Bose-Einstein condensates (BECs) provide one of the most active and creative areas of 
research in physics \cite{legget, dalfovo}. The dynamics of the BECs are usually described by 
a nonlinear Schr\"odinger equation (known in this field as the Gross-Pitaevskii equation (GPE) \cite{gross,pitaev}), 
which determines the evolution of a collective wave function of ultra-cold atoms in the BECs, 
evolving in the mean field self-potential.  

In this paper, we propose the use of an alternative but nearly equivalent approach to the physics of BECs, 
based on a kinetic equation for the condensate. We also show that this kinetic theory can lead to 
a more complete understanding of the physical processes occurring in the BECs, not only by 
providing an alternative method for describing the system, but also by improving our 
global view of the physical phenomena. It is our hope that this will also lead to the 
discovery of new aspects of BECs.

The key point of our present approach is the use of a Wigner Moyal equation (WME) for the BECs, 
describing the spatio-temporal evolution of the appropriate Wigner function \cite{hilary}. Wigner 
functions for the BECs were discussed in the past \cite{steel,gardiner} and the WME has been sporadically used \cite{gardiner2}. But no systematic application of the WME to BECs has previously been considered. In the quasi-classical limit, this equation reduces to 
the particle number conservation equation, which is a kinetic equation formally analogous to 
the Liouville equation, but with a nonlinear potential. A description of the BECs in terms of the 
kinetic equation is adequate to deal with a series of problems, as exemplified here, and can be 
seen as an intermediate (in accuracy) between the GPE and the 
hydrodynamic equations usually found in the literature. 

The manuscript is organized in the following fashion.  In Section 2, we establish the WME 
and discuss its approximate version as a kinetic equation for the Wigner function. We then apply the 
kinetic equation to two distinct physical problems. The first one, considered in Section 3, is the 
self-phase modulation of a BEC beam. A similar problem has been studied numerically in the past \cite{zhang}. 
Here,  we derive explicit analytical results and show that a part of the BEC beam is decelerated and 
eventually comes to a complete halt as a result of the collective forces acting on the condensate. 
The second example is considered in Section 4, where we establish a kinetic dispersion relation 
for sound waves in the BECs, giving a kinetic correction to the usual Bogoliubov sound 
speed \cite{andrews,zaremba} and predicting the occurrence of Landau damping \cite{lev, jackson}. 
Our description of Landau damping is  significantly different from that previously considered for 
transverse oscillations of  BECs \cite{guilleumas}. Finally, in Section 5, the virtues and limitations of 
the present kinetic approach are briefly discussed.

\section{Wigner Moyal Equation for the Bose Condensate}

It is known that for an ultra-cold atomic ensemble, and in particular for BECs, the ground state 
atomic quantum field can be replaced by a macroscopic atomic wave function $\psi$. In a large variety 
of situations, the evolution of $\psi$ is determined by the GPE 

\begin{equation}
i \hbar \frac{\partial \psi}{\partial t} = - \frac{\hbar^2}{2 m} \nabla^2 \psi 
+ (V_0 + V_{eff}) \psi, 
\label{eq:2.1} \end{equation}
where $V_0 \equiv V_0 (\vec{r})$ is the confining potential and $V_{eff}$ is the effective potential which 
takes into account the inter-atomic interactions inside the condensate, as determined in its simplest 
form by $V_{eff} (\vec{r}, t) = g |\psi (\vec{r}, t)|^2$, where $g$ is a constant \cite{gross,pitaev}.

Let us consider the situation where this wave equation can be replaced by a kinetic equation. 
In order to construct such an equation, we introduce the Wigner function associated with $\psi$, 
such that \cite{hilary}

\begin{equation}
W (\vec{r}, \vec{k}, t ) = \int \psi (\vec{r} + \vec{s}/2, t) \psi^*(\vec{r} 
- \vec{s}/2, t) \exp (- \vec{k} \cdot \vec{s} ) \; d \vec{s}.
\label{eq:2.2} \end{equation}

It is then possible to derive (see the Appendix) the following evolution equation for the Wigner function

\begin{equation}
\left( \frac{\hbar^2}{2 m} \vec{k} \cdot \nabla - i \hbar \frac{\partial}{\partial t} \right) W 
= - 2 V \sin \Lambda \; W,
\label{eq:2.3} \end{equation}
where a bi-directional differential operator is denoted by

\begin{equation}
\Lambda = \leftarrow \left( \frac{\partial}{\partial \vec{r}} \cdot  \frac{\partial}
{\partial \vec{p}} \right) \rightarrow.
\label{eq:2.4} \end{equation}
This operator acts to the left on $V$ and to the right on $W$ \cite{hilary}. 
In this equation, the potential is 

\begin{equation}
V = V_0 + g \int W (\vec{r}, \vec{k}, t) \frac{d \vec{k}}{(2 \pi)^3} + \delta V, 
\label{eq:2.5} \end{equation}
where 

\begin{equation}
\delta V = g \left( | \psi (\vec{r}, t) |^2 - \int W (\vec{r}, \vec{k}, t) \frac{d \vec{k}}{(2 \pi)^3} \right)
\label{eq:2.5b} \end{equation} 
can be considered a noise term associated with the square mean deviations of the quasi probability, 
determined by the Wigner function $W$ with respect to the local quantum probability, 
determined by the wave function $\psi$.

Equation (\ref{eq:2.3}) can be seen as the WME describing the space and time evolution 
of the BECs, and it is exactly equivalent to the GPE (\ref{eq:2.1}). However, it 
is of little use in the above exact form, and it is convenient to introduce some simplifying assumptions. 
This is justified for the important case of slowly varying potentials. In this case, we can neglect the 
higher order space derivatives, and introduce the approximation $\sin \Lambda \sim \Lambda$. This 
corresponds to the quasi-classical approximation, where the quantum potential fluctuations can 
also be neglected, viz. $\delta V \rightarrow 0$. Introducing these two simplifying assumptions, 
valid in the quasi-classical limit, we reduce the WME to a much simpler form

\begin{equation}
\left( \frac{\partial}{\partial t} + \vec{v} \cdot \nabla + \vec{F} \cdot \frac{\partial}
{\partial \vec{k}} \right) W = 0, 
\label{eq:2.6} \end{equation}
where $\vec{v} = \hbar \vec{k}/m$ is the velocity of the condensate atoms corresponding to the 
wavevector state $\vec{k}$, and $\vec{F} = \nabla V$ is a force associated with the inhomogeneity 
of the condensate self-potential. The nonlinear term in the GPE (\ref{eq:2.1}) 
is hidden inside this force $\vec{F}$. As we will see, this nonlinear term will look very much 
like a ponderomotive force term, similar to the radiation pressure.

It should be noticed that this new equation is a closed kinetic equation for the Wigner function $W$. 
In this quasi-classical limit, $W$ is just the particle occupation number for translational states 
with momentum $\vec{p} = \hbar \vec{k}$. Equation (\ref{eq:2.6}) is equivalent to a conservation equation, 
stating the conservation of the quasi-probability $W$ in the six-dimensional classical phase 
space $(\vec{r}, \vec{k})$, and can also be written as

\begin{equation}
\frac{d}{d t} W (\vec{r}, \vec{k}, t) = 0.
\label{eq:2.6b} \end{equation}

This kinetic equation can then be used to describe physical processes occurring in a BEC, as long as 
the quasi-classical approximation of slowly varying potentials is justified. The interest of such a 
kinetic descriptions will be illustrated with the aid of two simple and different examples, to be 
presented in the next two sections. Many other applications can be envisaged, and will be 
explored in future publications.

\section{Self-phase modulation of a beam condensate}

Let us first consider the kinetic description of self-phase modulation of a BEC gas, moving with respect 
to the confining potential $V_0 (\vec{r})$.  Here, we can explore the analogy of this problem with that 
of self-phase modulation of short laser pulses moving in a nonlinear optical medium and well known in the 
literature \cite{silva}. In order to simplify our description, we consider the one-dimensional problem 
of a beam moving along the $z$-axis and neglect the axial variation of the background potential 
$\partial V_0 / \partial z \sim 0$. The radial structure of the beam can easily be introduced later, 
and will not essentially modify the results obtained here. The kinetic equation (\ref{eq:2.6}) 
can then be written as

\begin{equation}
\left( \frac{\partial}{\partial t} + v_z  \frac{\partial}{\partial z} 
+ F_z  \frac{\partial}{\partial k} \right) W (z, k, t) = 0,
\label{eq:3.1} \end{equation}
with $v_z$ and $F_z$ given by, respectively,

\begin{equation}
v_z = \frac{\hbar k}{m} + g  \frac{\partial}{\partial t} I (z, t) \quad , \quad
F_z = \frac{d k}{d t} = - g  \frac{\partial}{\partial z} I (z, t),
\label{eq:3.2} \end{equation}
where we have used the intensity of the beam condensate, as defined by

\begin{equation}
I (z, t) = \int W (z, k, t) \frac{d k}{2 \pi}.
\label{eq:3.2b} \end{equation}

Let us assume that the ultra-cold atomic beam has a mean velocity $v_0 = \hbar k_0/m$. This suggests 
the use of a new space coordinate $\eta = z - v_0 t$. In terms of this new coordinate, the quasi-classical 
equations of motion of a cold atom in the beam can be written as

\begin{eqnarray}
\frac{d \eta}{d t} &=& \frac{\partial h}{\partial k} = \frac{1}{m}  (k - k_0), 
\\ \nonumber
\frac{d k}{d t} &=& - \frac{\partial h}{\partial \eta} 
= - \frac{g}{\hbar} \frac{\partial}{\partial \eta}  I (\eta, t),
\label{eq:3.3} \end{eqnarray}
where we have introduced the Hamiltonian function

\begin{equation}
h (\eta, k, t) = \omega (\eta, k, t)  - k v_0 = \frac{k}{m} \left( \frac{k}{2} - k_0 \right) 
+ \frac{g}{\hbar}  I (\eta, t).
\label{eq:3.4} \end{equation}
Here $\omega (\eta, k, t)$ is the Hamiltonian in the rest frame expressed in the new coordinate. 
A straightforward integration of the equations of motion leads to

\begin{equation}
k (t) = k_0 -  \frac{g}{\hbar} \int_0^t  \frac{\partial}{\partial \eta}  I (\eta, t') dt'.
\label{eq:3.5} \end{equation}

At this point it is useful to introduce the concept of the beam energy chirp, $< \epsilon (\eta, t)>$, in 
analogy with the frequency chirp of short laser pulses \cite{silva}. By definition, it will be the 
beam mean energy at a given position and at a given time

\begin{equation}
< \epsilon (\eta, t) > = \hbar \int W (\eta, k, t) \omega (\eta, k, t) \frac{dk}{2 \pi},
\label{eq:3.6} \end{equation}
where the weighting function $W (\eta, k, t)$ is the solution of the kinetic one-dimensional 
equation (\ref{eq:3.1}). A formal solution of this equation can be written as

\begin{equation}
W (\eta, k, t) = W (\eta_0 (\eta, k, t), k_0 (\eta, k, t), t_0),
\label{eq:3.7} \end{equation}
where $\eta_0$ and $k_0$ are the initial conditions corresponding to the observed values at time $t$, 
as determined by the dynamical equations (\ref{eq:3.3}). Replacing it in equation (\ref{eq:3.5}), we obtain

\begin{equation}
< \epsilon (\eta, t) > = \hbar \int W (\eta_0, k_0, t_0) 
\left[  \frac{k^2}{2m} + \frac{g}{\hbar}  I (\eta, t) \right] \frac{dk}{2 \pi}.
\label{eq:3.8} \end{equation}

From equation (\ref{eq:3.5}) we notice that $d k = d k_0$. Neglecting higher order nonlinearities, 
we can then rewrite the above expression as

\begin{equation}
< \epsilon (\eta, t) > = < \epsilon (0) > - \frac{k_0}{m} g \int_0^t  \frac{\partial}{\partial \eta}  
I (\eta, t') dt' ,
\label{eq:3.9} \end{equation}
where $< \epsilon (0) > \equiv < \epsilon (\eta_0, t_0) >$ is the initial beam energy chirp.

Let us first consider that the beam profile $I (\eta)$ is independent of time. This is, of course, 
only valid for very short time intervals where the beam velocity dispersion is negligible. 
In this simple case, we have

\begin{equation}
< \epsilon (\eta, t) > = < \epsilon (0) > - \hbar v_0 g \left( \frac{\partial I}{\partial \eta} \right) t.
\label{eq:3.9a} \end{equation}

The maximum energy shift will be attained at some position inside the beam profile,  
$\eta = \eta_{max}$, determined by the stationarity condition

\begin{equation}
\frac{\partial}{\partial \eta} < \epsilon (\eta, t) > = \left( \frac{\partial^2 I}{\partial \eta^2} \right) = 0.
\label{eq:3.9b} \end{equation}

In order to deduce more specific answers, let us assume a Gaussian beam profile

\begin{equation}
I (\eta) = I_0 \exp \left( - \eta^2 / \sigma^2 \right),
\label{eq:3.10} \end{equation}
where $\sigma$ determines the beam width. For this profile we have $\eta_{max} = \pm \sigma / \sqrt{2}$, 
which leads to the following value of the maximum energy shift

\begin{equation}
\Delta \epsilon (t) \equiv < \epsilon (t) >_{max} - < \epsilon (0) > 
= \pm \frac{\hbar \sqrt{2}}{\sigma} g v_0 I_0 e^{-1/2} t.
\label{eq:3.11} \end{equation}
This is similar to a well known result in nonlinear optics, stating that the maximum energy chirp 
due to a self-phase modulation is proportional to $t$, or to the distance travelled by the beam, 
$d = v_0 t$. This result clearly indicates that the initial beam will eventually split into two parts, 
one being accelerated to higher translational speeds, and the other being decelerated. This would correspond 
to the red-shift and to the blue-shift observed in nonlinear optics. The decelerated beam will eventually 
stop after a time $t \simeq \tau$, such that $\Delta \epsilon (\tau) = < \epsilon (0) >$. This will 
determine the condition for translational beam freezing.

It should be noticed that the same result could also be obtained directly from the GPE (\ref{eq:2.1}). 
But the interest of the present derivation is that it demonstrates the 
irrelevance of the phase of the wave function $\psi$, because such a phase was ignored in our kinetic 
calculation. Therefore, instead of the self-phase modulation, it would be more appropriate to call 
it a beam self-deceleration. 

Another interesting aspect of our kinetic approach is that it can be easily refined, as shown briefly here. 
Let us then improve the above calculation by considering the beam dispersion. This will inevitably occur 
because of the linear velocity dispersion of the atomic beam. Such a dispersion will decrease the chirping 
effect, because of the decrease in time of $(\partial I / \partial \eta)$. In order to model it, we can 
assume a time-varying Gaussian beam shape, as described by

\begin{equation}
I (\eta, t) = I_0 \left( \frac{\sigma_0}{\sigma (t)} \right)^{1/2} 
\exp \left( - \frac{\eta^2 }{ \sigma (t) ^2} \right).
\label{eq:3.12} \end{equation}
If we now consider that $\sigma (t) = \sigma_0 (1 + \delta t^2)$, where $\delta 
= (2 m/\hbar^2) \Delta \epsilon_0 / \sigma_0$ is proportional to the initial energy spread 
$\Delta \epsilon_0$. This will lead to a maximum energy shift 

\begin{equation}
\Delta \epsilon_d (t) = \frac{\ln (t)}{t} \Delta \epsilon (t). 
\label{eq:3.14} \end{equation}
It is clear that  the linear beam velocity dispersion will decrease the maximum attainable chirp, by 
changing the linearity with time into a logarithmic law. However, this will only occur for very long times, 
$t \sim 1/ \sqrt{\delta}$, which for ultra-cold atomic beams with a very low translational energy 
dispersion $\Delta \epsilon_0$ will not be relevant.

The other cause of the beam dispersion is the nonlinear process itself, which will eventually break up the 
initial pulse into two distinct pulses. In this case, the self-phase modulation process will not be attenuated 
because the beam width will be conserved, but the two secondary pulses will suffer self-phase modulation 
themselves, and will eventually break up later, resulting in the formation of several secondary pulses 
with different mean energy. However, the nonlinear dispersion will also be negligible as far as 
$\sigma_0^2 > 4 m  | \Delta \epsilon (t) | t^2 / \hbar^2$. A more complete description of all these 
dispersion regimes can be obtained by solving numerically the kinetic equation (\ref{eq:3.1}).

\section{Kinetic description of Bogoliubov oscillations}

The second example of an application of the kinetic equation for the BECs deals  with the dispersion relation 
of sound waves. For simplicity, we consider again the one-dimensional model and neglect the radial structure 
of the oscillations. This allows us to treat the lowest order oscillating modes of the condensate. We assume 
some given equilibrium distribution $W_0 (z, k, t)$, corresponding, for instance, to the Thomas-Fermi 
equilibrium solution in a given confining potential $V_0 (\vec{r}_\perp, z)$ \cite{byam}, and after 
linearization of the one-dimensional kinetic equation (\ref{eq:3.1}) with respect to the perturbation $\tilde{W}$, 
we obtain 

\begin{equation}
\left( \frac{\partial}{\partial t} + v_z  \frac{\partial}{\partial z} \right) \tilde{W} (z, k, t)
+ \tilde{F}  \frac{\partial}{\partial k} W_0 (z, k, t) = 0,
\label{4.1} \end{equation}
where the perturbed force is determined by

\begin{equation}
\tilde{F} = -  \frac{g}{\hbar} \frac{\partial}{\partial z} \tilde{I} (z, t) 
= -  \frac{g}{\hbar} \frac{\partial}{\partial z} \int \tilde{W} (z, k, t) \frac{d k}{2 \pi}.
\label{eq:4.2} \end{equation}

Let us now assume perturbations of the form $\tilde{W}, \tilde{I} \sim \exp(i k z - i \omega t)$. From the 
above equations we then obtain a relation between the perturbation amplitude of the Wigner function $\tilde{W}$ 
and the perturbed beam intensity $\tilde{I}$

\begin{equation}
\tilde{W} = - \frac{g k}{\hbar (\omega - k v')} \tilde{I}  \frac{\partial}{\partial k'} W_0 (k'),
\label{eq:4.3} \end{equation}
where we now specify the particle wavenumber state with $k'$, in order to avoid confusion with the wavenumber 
$k$ of the oscillation that we intend to study. The velocity corresponding to this particle state is 
$v' = \hbar k'/m$. Integration over the momentum spectrum of the particle condensate will then lead 
to the following expression

\begin{equation}
1 + \frac{g}{\hbar} k \int \frac{\partial W_0 (k') / \partial k'}
{(\omega - \hbar k k' / m)} \frac{d k'}{2 \pi} = 0.
\label{eq:4.4} \end{equation}
This is the kinetic dispersion relation for axial perturbations in the BECs. Let us illustrate this result 
by considering a simple case for a condensate beam with no translational dispersion, or with a translational 
temperature exactly equal to zero. The equilibrium state of the beam can then be described by

\begin{equation}
W_0 (k') = 2 \pi n_0 \delta(k' - k'_0),
\label{eq:4.5} \end{equation}
where $n_0 = \int W_0 (k') d k' / 2 \pi$ is the particle number density in the condensate. 
Replacing this in the dispersion relation (\ref{eq:4.4}). we have

\begin{equation}
1 - \frac{g k^2}{m} \frac{n_0}{(\omega - k v'_0)^2} = 0,
\label{eq:4.6} \end{equation}
where $v'_0 = \hbar k'_0 / m = p'_0 / m$ is the beam velocity. This can also be written as

\begin{equation}
(\omega - k v'_0)^2 = k^2 c_s^2,
\label{eq:4.7} \end{equation}
where

\begin{equation}
c_s = \sqrt{g n_0 / m}
\label{eq:4.7b} \end{equation}
is nothing but the Bogoliubov sound speed. Obviously, equation (\ref{eq:4.7}) is the Doppler shifted 
dispersion relation of sound waves in the BEC gas, In its reference frame it reduces to $\omega = k c_s$.

Let us now consider a situation where, instead of the distribution (\ref{eq:4.5}), we have a beam with 
a small translational velocity spread, such that the number of particles with a velocity $v' \sim c_s$ is 
small but nonzero. In this case, the resonant contribution in the integral of equation (\ref{eq:4.4}) has 
to be retained, although it is still possible to neglect the kinetic corrections in the principal part of 
the integral. The dispersion relation can then be written, in the condensate frame of reference, as

\begin{equation}
1 - \frac{k^2 c_s^2}{\omega^2} - \frac{i}{2} \frac{g m}{\hbar^2} 
\left( \frac{\partial W_0}{\partial k'} \right)_{k' = k'_s} = 0,
\label{eq:4.8} \end{equation}
where $k'_s = m c_s / \hbar$ is the resonant momentum. The imaginary term in this equation can then lead 
to damping of the sound waves. Writing now $\omega = k c_s + i \gamma$, with $|\gamma| \ll k c_s$, we 
can then obtain the expression for the damping coefficient

\begin{equation}
\gamma = \frac{\omega}{4} \frac{g m}{\hbar^2} \left( \frac{\partial W_0}{\partial k'} \right)_{k' = k'_s}. 
\label{eq:4.9} \end{equation}
The above expression corresponds to the non-collisional Landau damping of Bogoliubov oscillations in the 
BECs. The present approach can also be generalized in a straightforward way to higher order oscillations 
of the condensate, where the radial structure has to be taken into account \cite{zaremba,string}.

\section{Conclusions}

In this paper, we have proposed a kinetic view of the Bose-Einstein condensate physics, based on the Wigner-Moyal equation. 
In the quasi-classical limit, the latter can be reduced to a closed kinetic equation for the corresponding 
Wigner function. The kinetic approach to BEC can be seen as an intermediate step between the GPE and 
the hydrodynamical equations for the condensate gas, often found in the literature. 

We have discussed two different physical problems, in order to illustrate the versatility 
of the kinetic theory. One is a self-phase modulation of a BEC beam. The other is the dispersion relation of 
the Bogoliubov oscillations in the condensate gas. The first example shows that, due to the influence of 
its own inhomogeneous self-potential, nearly half of the beam is accelerated while the other half is 
decelerated. Under certain conditions, the decelerated part of the beam will tend to a state of complete halt. 
The second example shows that a kinetic dispersion relation for sound waves in  BECs can be established, 
where Landau damping is automatically included.  The present results only considered the lowest order modes, 
but the same approach can be used to describe higher order oscillations of the BECs, including  their 
radial structures, as well as the coupling with a background thermal gas. This investigation is,
however, beyond the scope of the present work.

Several other different problems relevant to BECs can also be considered in the frame of the kinetic theory, 
such as modulational instabilities \cite{konotop} and the wakefield generation. This indicates that the kinetic theory is a 
very promising approach to the physics of BECs, which will eventually allow us to introduce new ideas in this 
stimulating area of research, and to suggest new configurations to the experimentalists. However, the present 
work also clearly states that the present theory is only valid in the quasi-classical limit and for that 
reason some relevant problems, where the phase of the BEC wave function plays an important role, can only be 
treated by means of the GPE. Surprisingly, the self-phase modulation is not one of them, as demonstrated here.

\acknowledgements 
{One of the authors (J.T.M.) appreciates the hospitality of the Rutherford 
Appleton Laboratory, and partial support of the Centre for Fundamental Physics.}

\newpage

\bigskip

{\bf APPENDIX: Derivation of the Wigner-Moyal equation}

In the present derivation we follow a procedure already used in other cases, for instance, in the case of 
electromagnetic waves moving in a space and time dependent dielectric medium \cite{nodar}. For a different but nearly equivalent derivation of the WME see the appendix of reference \cite{gardiner2}. Let us consider 
two distinct sets of values for space and time coordinates, $(\vec{r}_1, t_1)$ and $(\vec{r}_2, t_2)$, 
and  let us use the notation $\psi_j = \psi (\vec{r}_j, t_j)$ and $V_j = V (\vec{r}_j, t_j)$, for $j = 1, 2$. 
This allows us to write two versions of the GPE (\ref{eq:2.1}) as

\begin{equation}
\left( \frac{\hbar^2}{2 m} \nabla^2_j - i \hbar \frac{\partial}{\partial t_j} \right) \psi_j = - V_j \psi_j.
\label{eq:a.1} 
\end{equation}

Multiplying the equation $j = 1$ by $\psi_2^*$ and the conjugate of the equation $j = 2$ by $\psi_1$, and 
subtract the resulting equations, we obtain

\begin{equation}
\left[ \frac{\hbar^2}{2 m} ( \nabla^2_1 - \nabla^2_2) - i \hbar \left( \frac{\partial}{\partial t_1} 
+ \frac{\partial}{\partial t_2} \right) \right] C_{12} = - (V_1 - V_2) C_{12},  
\label{eq:a.2} \end{equation}
where we have used $C_{12} = \psi_1 \psi_2^*$. The above equation suggests the use of two pairs of 
space and time variables, such that

\begin{eqnarray}
\vec{r}_1 = \vec{r} - \vec{s} / 2 \quad , \quad t_1 = t - \tau / 2,
\\ \nonumber
\vec{r}_2 = \vec{r} + \vec{s} / 2 \quad , \quad t_2 = t + \tau / 2.
\label{eq:a.3} \end{eqnarray}
We can then rewrite the above equation as

\begin{equation}
\left[ \frac{\hbar^2}{m} \frac{\partial}{\partial \vec{r}} \cdot  
\frac{\partial}{\partial \vec{s}} - i \hbar  \frac{\partial}{\partial t}  \right] C_{12} 
= - (V_1 - V_2) C_{12}.
\label{eq:a.4} \end{equation}

It can also easily be shown, by developing the potentials $V_j$ around $V    (\vec{r}, t)$, that

\begin{equation}
(V_1 - V_2) = 2 \sinh \left( \frac{\vec{s}}{2} \cdot \frac{\partial}{\partial \vec{r}} 
+ \frac{\tau}{2} \frac{\partial}{\partial t} \right) V (\vec{r}, t).
\label{eq:a.5} \end{equation}

Let us now introduce the double Fourier transformation of the function $C_{12} \equiv 
C(\vec{r}, \vec{s}, t, \tau) $ on the variables $\vec{s}$ and $\tau$, as defined by

\begin{equation}
W (\vec{r}, t, \omega, \vec{k}) = \int d \vec{s} \int d \tau C(\vec{r}, \vec{s}, t, \tau) 
\exp (- i \vec{k} \cdot \vec{s} + i \omega \tau). 
\label{eq:a.6} \end{equation}
This can be rewritten in terms of the wave function $\psi$ as

\begin{equation}
W (\vec{r}, t, \omega, \vec{k}) = \int d \vec{s} \int d \tau \psi 
(\vec{r} + \vec{s} / 2, t + \tau/ 2) \psi^* (\vec{r} - \vec{s} / 2, t - \tau/ 2)
\exp (- i \vec{k} \cdot \vec{s} + i \omega \tau). 
\label{eq:a.6b} \end{equation}

Using such a development in equation (\ref{eq:a.4}), we obtain for the Fourier amplitudes

\begin{equation}
\left( \frac{\hbar^2}{m} \vec{k} \cdot \frac{\partial}{\partial \vec{r}} 
- \hbar \frac{\partial}{\partial t} \right) W = - 2 V \sin \Lambda' W,
\label{eq:a.7} \end{equation}
where we have used the following differential operator, operating backwards on the 
potential $V (\vec{r}, t)$ and forward on $W$, viz.

\begin{equation}
\Lambda' = \frac{1}{2} ^\leftarrow \left( \frac{\partial}{\partial \vec{r}} \cdot 
\frac{\partial}{\partial \vec{k}} - \frac{\partial}{\partial t} 
\frac{\partial}{\partial \omega} \right)^\rightarrow . 
\label{eq:a.7b} \end{equation}

This is a formidable equation for $W$, which can be simplified by noting that the GPE 
implies the existence of a well defined relation between energy and momentum. This means that $\omega$ must 
be equal to some function of $\vec{k}$, or $\omega = \omega (\vec{k})$. Hence, we can state that

\begin{equation}
W (\vec{r}, t, \omega, \vec{k}) = W (\vec{r}, \vec{k}, t) \delta (\omega - \omega (\vec{k})). 
\label{eq:a.8} \end{equation}

This leads to a much simpler evolution expression for $W (\vec{r}, \vec{k}, t)$. Before writing it down, 
we should also notice that the nonlinear term in $V$ depends on $|\psi|^2$, and not on the function $W$. 
Thus, we can finally write

\begin{equation}
\left( \frac{\hbar^2}{2 m} \vec{k} \cdot \nabla - i \hbar \frac{\partial}{\partial t} \right) W 
= - 2 (V_0 + g |\psi|^2)  (\sin \Lambda) W,
\label{eq:a.9} \end{equation}
where $\Lambda$ is a simpler differential operator defined by

\begin{equation}
\Lambda = \; \leftarrow \left( \frac{\partial}{\partial \vec{r}} \cdot  
\frac{\partial}{\partial \vec{p}} \right) \rightarrow .
\label{eq:a.9b} \end{equation}

The function $W (\vec{r}, \vec{k}, t) $ can be seen as the Wigner function associated with the GPE, 
and equation (\ref{eq:a.9}) as the WME equation that describes its spatio-temporal behavior. 
This equation is equivalent to the initial wave equation (\ref{eq:2.1}), but it is not a closed equation for 
the quasi-probability function $W$. Therefore, some simplifying assumptions have to be introduced in order 
to make it more tractable, as explained in Section 2.

\end{document}